\documentclass[a4paper,10pt]{article}

\usepackage{bm,amsmath,amssymb,graphicx,mathrsfs}

\newcommand{\nc}{\newcommand}
\nc{\rnc}{\renewcommand}
\nc{\nn}{\nonumber}
\nc{\bra}{\langle}
\nc{\ket}{\rangle}
\rnc{\(}{\left(}
\rnc{\)}{\right)}
\rnc{\[}{\left[}
\rnc{\]}{\right]}

\nc{\sh}{\mathrm{sh}}

\begin{document}

\title{Two point functions for the six vertex model
with reflecting end}

\author{Kohei Motegi\thanks{E-mail: motegi@gokutan.c.u-tokyo.ac.jp} \\
\it Okayama Institute for Quantum Physics, \\
\it Kyoyama 1-9-1,
Okayama 700-0015, Japan}

\date{\today}

\maketitle

\begin{abstract}
The two point functions,
which give the probability that
the spins turn down at the boundaries, 
are studied for the six vertex model
on a $2N \times N$ lattice with
domain wall boundary condition and left reflecting end.
We consider two types of two point functions,
and express them using determinants.
\end{abstract}

\section{Introduction}
The six vertex model is 
one of the most fundamental exactly solved models
in statistical physics \cite{Slater,Lieb,Sutherland,Baxter}.
Not only the periodic boundary condition
but also the domain wall boundary condition
is an interesting boundary condition.
For example, the partition function is deeply 
related to the norm \cite{Korepin} and the scalar product
\cite{Slavnov} of the XXZ chain.
The determinant formula of the partition function \cite{Izergin,ICK}
lead Slavnov \cite{Slavnov} to obtain a compact representation of
the scalar product, which plays a fundamental role
in calculating correlation functions of the XXZ chain
\cite{KIEU,KIB,KMT,KMST}.
The determinant formula also led to a deep advance
in enumerative combinatorics
\cite{Zeilberger,Kuperberg,Bressoud}.
For example, it was used to give a concise proof of the 
numbers of the alternating sign matrices
for a given size.
Recently, the correspondences between the partition function
and the Schur polynomial \cite{La} and KP $\tau$ function
\cite{FWZ} have been revealed.

The calculation of correlation functions are also
interesting in the domain wall boundary condition itself.
Several kinds of them such as the boundary correlation functions
\cite{BKZ,BPZ,FP,CP}
and the emptiness formation probability \cite{CP2} have been calculated.
Some of them are shown to be expressed in determinant forms.

The mixed boundary conditions of the domain wall and reflecting
boundary \cite{Sklyanin} has also been studied.
The partition function \cite{Tsuchiya}
and several kinds of boundary one point functions \cite{Wang,Motegi}
are computed and expressed using determinants.

In this paper, we calculate
boundary two point functions for the six vertex model
on a $2N \times N$ lattice with the mixed boundary condition.
The two point functions we consider give the probability that
the spins turn down at the boundaries.
We calculate two types of two point functions,
and express them in terms of determinants.

The outline of this paper is as follows.
In the next section, we define the six vertex model
with mixed boundary condition.
We consider two types of two point functions,
and call them Type $\mathrm{I}$ and Type $\mathrm{II}$,
according to the boundaries the spins we consider
are associated with.
Type $\mathrm{I}$ is evaluated in section 3, 
Type $\mathrm{II}$ in section 4.
\section{Six vertex model}
\begin{figure}[htbp]
 \begin{center}
  \includegraphics[width=100mm]{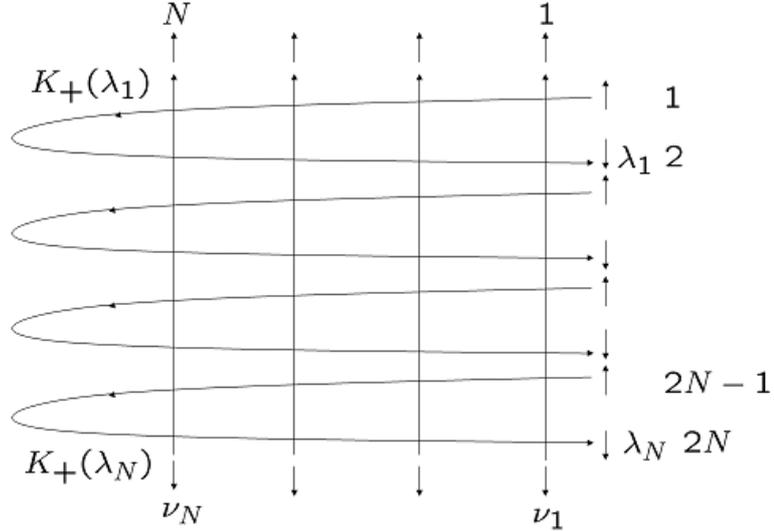}
 \end{center}
 \caption{The six vertex model with left reflecting boundary.}
 \label{fig:one}
\end{figure}

The six vertex model is a model in statistical mechanics,
whose local states
are associated with edges of a square lattice,
which can take two values.
The Boltzmann weights are assigned to its vertices,
and each weight is determined by the configuration
around a vertex. What plays the fundamental role is the $R$-matrix
\begin{align}
R(\lambda)
=&\left(
\begin{array}{cccc}
a(\lambda)  & 0 & 0 & 0 \\
0 & b(\lambda) & c(\lambda) & 0 \\
0 & c(\lambda) & b(\lambda) & 0 \\
0 & 0 & 0 & a(\lambda)
\end{array}
\right),
\end{align}
where
\begin{align}
a(\lambda)=1, \ b(\lambda)=\frac{ \sh \lambda}{ \sh (\lambda+\eta)},
\ c(\lambda)=\frac{ \sh \eta}{ \sh (\lambda+\eta)}.
\end{align}
The $R$-matrix satisfies the Yang-Baxter equation
\begin{align}
R_{12}(\lambda) R_{13}(\lambda+\mu) R_{23}(\mu)
=R_{23}(\mu) R_{13}(\lambda+\mu) R_{12}(\lambda).
\label{YangBaxter}
\end{align}
In this paper, we consider the six vertex model
on a $2N \times N$ lattice depicted in Figure \ref{fig:one}.
At the upper and lower boundaries, the spins are aligned
all up and all down respectively.
At the right boundary, 
the boundary spins are up for odd rows and down for even rows.
We set $L_{\alpha k}(\lambda_{\alpha},\nu_k)
=R_{\alpha k}(\lambda_{\alpha}-\nu_k-\eta/2)$.
At the intersection of the $\alpha$-th row
and the $k$-th column, we associate the statistical weight
$\sigma_{\beta}^2 L_{\beta k}(-\lambda_{\beta}, \nu_k) \sigma_{\beta}^2,
\ \beta=(\alpha+1)/2$
for $\alpha$ odd and $L_{\beta k}^{t_{\beta}}(\lambda_{\beta},\nu_k),
\ \beta=\alpha/2$
for $\alpha$ even.
Between the $(2 \alpha-1)$-th and $(2 \alpha)$-th row,
the boundary statistical weight
\begin{align}
K_+(\lambda_{\alpha})
=&\left(
\begin{array}{cc}
\sh(\lambda_{\alpha}+\eta/2+\zeta_{+})  & 0 \\
0 & \sh(-\lambda_{\alpha}-\eta/2+\zeta_{+})
\end{array}
\right), \label{Kmatrix}
\end{align}
is associated at the left boundary.
For later convenience, we denote 
$\{ \lambda \}=\{\lambda_1, \lambda_2, \dots , \lambda_N  \},
\{ \nu \}=\{\nu_1, \nu_2, \dots , \nu_N  \}$,
and introduce the one-row monodromy matrix
\begin{align}
T(\lambda_{\alpha}, \{ \nu \})=&L_{\alpha N}(\lambda_{\alpha},\nu_N)
\cdots L_{\alpha 1}(\lambda_{\alpha},\nu_1) \nonumber \\
=&\left(
\begin{array}{cc}
A(\lambda_{\alpha}, \{ \nu \})  & B(\lambda_{\alpha}, \{ \nu \}) \\
C(\lambda_{\alpha}, \{ \nu \}) & D(\lambda_{\alpha}, \{ \nu \})
\end{array}
\right). \label{onerow}
\end{align}
Combining two one-row monodromy matrices
and the $K$-matrix \eqref{Kmatrix},
one can construct the double-row monodromy matrix
\begin{align}
U^{t_{\alpha}}(\lambda_{\alpha}, \{ \nu \})&=
T^{t_{\alpha}}(\lambda_{\alpha}, \{ \nu \})K_+(\lambda_{\alpha})
\sigma_{\alpha}^2 T(-\lambda_{\alpha}, \{\nu \}) \sigma_{\alpha}^2,
\nonumber \\
&=\left(
\begin{array}{cc}
\mathcal{A}(\lambda_{\alpha}, \{ \nu \})  & \mathcal{C}(\lambda_{\alpha}, 
\{ \nu \}) \\
\mathcal{B}(\lambda_{\alpha}, \{ \nu \}) & \mathcal{D}(\lambda_{\alpha},
\{ \nu \})
\end{array}
\right).
\end{align}
The partition function of the six vertex model
with mixed boundary condition, which is the summation
of products of statistical weights over all possible configurations, can
be represented as
\begin{align}
Z_{2N \times N}(\{ \lambda \}, \{ \nu \})
=w_N^- \mathcal{B}(\lambda_{N}, \{ \nu \}) \cdots
\mathcal{B}(\lambda_{1}, \{ \nu \})  w_N^+,
\end{align}
where $w_{N}^+= \prod_{k=1}^{N} \uparrow_k$ and
$w_{N}^-= \prod_{k=1}^{N} \downarrow_k$. It
has the following determinant form \cite{Tsuchiya}
\begin{align}
Z_{2N \times N}(\{ \lambda \}, \{ \nu \})
=\frac{\prod_{j=1}^N \prod_{k=1}^N 
\[ \sh^2(\nu_j+\eta/2)-\sh^2 \lambda_k \]
\det_N \chi(\{ \lambda \}, \{ \nu \})
}
{
\prod_{1 \le j < k \le N}\[ \sh^2 \nu_j -\sh^2 \nu_k \]
\prod_{1 \le j < k \le N}\[ \sh^2 \lambda_k -\sh^2 \lambda_j \]
},
\label{boundarypartition}
\end{align}
where $\chi$ is an $N \times N$ matrix whose elements are given by
\begin{align}
\chi_{jk}&=\chi(\lambda_j, \nu_k), \\
\chi(\lambda, \nu)
&=\frac{-\sh \eta \sh (2 \lambda+\eta) \sh(\nu+\zeta_+)}
{\[ \sh^2(\nu+\eta/2)-\sh^2 \lambda \]
\[  \sh^2(\nu-\eta/2)-\sh^2 \lambda \]}.
\end{align}
\section{Type $\mathrm{I}$}
\begin{figure}[htbp]
 \begin{center}
  \includegraphics[width=100mm]{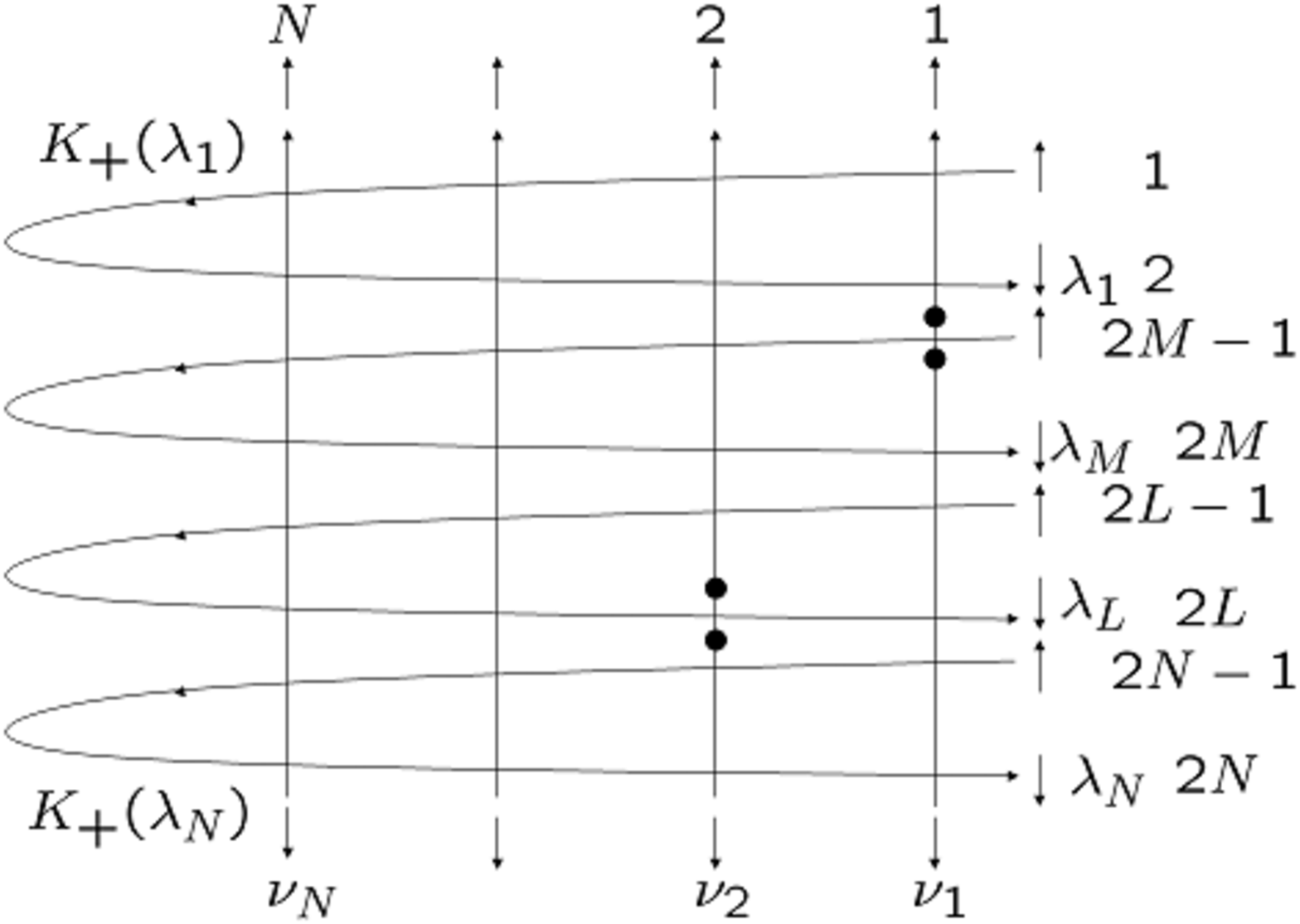}
 \end{center}
 \caption{Type $\mathrm{I}$ \eqref{type1}.}
 \label{fig:type1}
\end{figure}

In this section, we consider the following
two point function which we call Type $\mathrm{I}$.
\begin{align}
\Psi_1(M,L)=\frac{\psi_1(M,L)}
{Z_{2N \times N}(\{ \lambda \}, \{ \nu \})},
\label{type1}
\end{align}
where 
\begin{align}
\psi_1(M,L)=&w_N^- \mathcal{B}(\lambda_{N}, \{ \nu \})
\cdots \mathcal{B}(\lambda_{L+1}, \{ \nu \})
\mathcal{G}_2(\lambda_{L}, \{ \nu \})
\mathcal{B}(\lambda_{L-1}, \{ \nu \}) \cdots
\mathcal{B}(\lambda_{M+1}, \{ \nu \}) \nonumber \\
&\times \mathcal{F}(\lambda_{M}, \{ \nu \})
\mathcal{B}(\lambda_{M-1}, \{ \nu \}) \cdots
\mathcal{B}(\lambda_{1}, \{ \nu \}) w_N^-,
\end{align}
\begin{align}
\mathcal{F}(\lambda_\alpha, \{ \nu \})&=\downarrow_{2 \alpha}
W(\lambda_\alpha, \{ \nu \}) \uparrow_{2 \alpha-1}, \\
W(\lambda_\alpha, \{ \nu \})&=T^{t_\alpha}(\lambda_\alpha, \{ \nu \})
K_+(\lambda_\alpha) \frac{1}{2}(1-\sigma_1^3)
\sigma_\alpha^2 T(-\lambda_\alpha, \{ \nu \}) \sigma_\alpha^2
\frac{1}{2}(1+\sigma_1^3), \\
\mathcal{G}_2(\lambda_\alpha, \{ \nu \})&=\downarrow_{2 \alpha}
X_2(\lambda_\alpha, \{ \nu \}) \uparrow_{2 \alpha-1}, \\
X_2(\lambda_\alpha, \{ \nu \})&=\frac{1}{2}(1-\sigma_2^3)
T^{t_\alpha}(\lambda_\alpha, \{ \nu \}) \frac{1}{2}(1+\sigma_2^3)
K_+(\lambda_\alpha)
\sigma_\alpha^2 T(-\lambda_\alpha, \{ \nu \}) \sigma_\alpha^2.
\end{align}
This two point function is depicted in Figure \ref{fig:type1},
and gives the probability that the spin
on the first column is turned down just
on the $(2M-1)$-th row, and the spin on the second column
is turned down just on the $(2L)$-th row.
We calculate this two point function in several steps,
and show that as in \cite{CP}, it can be expressed utilizing a determinant
whose entries include differential operators. \\
First, from the graphical representation, we find
the numerator $\psi_1(M,L)$ becomes
\begin{align}
\psi_1(M,L)=&-c(-\lambda_M-\nu_1-\eta/2)
\prod_{j=1}^{M-1} \{b(-\lambda_j-\nu_1-\eta/2)
b(\lambda_j-\nu_1-\eta/2)  \} \nonumber \\
&\times w_{N-1}^-
\mathcal{B}(\lambda_N, \{ \nu \} \backslash \nu_1)
\cdots \mathcal{B}(\lambda_{L+1}, \{ \nu \} \backslash \nu_1)
\mathcal{G}(\lambda_L, \{ \nu \} \backslash \nu_1)
\nonumber \\
&\times \mathcal{B}(\lambda_{L-1}, \{ \nu \} \backslash \nu_1)
\cdots
\mathcal{B}(\lambda_{M+1}, \{ \nu \} \backslash \nu_1)
\mathcal{D}(\lambda_M, \{ \nu \} \backslash \nu_1)
\nonumber \\
&\times \mathcal{B}(\lambda_{M-1}, \{ \nu \} \backslash \nu_1)
\cdots
\mathcal{B}(\lambda_1, \{ \nu \} \backslash \nu_1)
w_{N-1}^+,
\end{align}
where
\begin{align}
\mathcal{G}(\lambda_\alpha, \{ \nu \} \backslash \nu_1)
&=\downarrow_{2 \alpha}
X(\lambda_\alpha, \{ \nu \} \backslash \nu_1) \uparrow_{2 \alpha-1}, \\
X(\lambda_\alpha, \{ \nu \} \backslash \nu_1)&=\frac{1}{2}(1-\sigma_2^3)
T^{t_\alpha}(\lambda_\alpha, \{ \nu \} \backslash \nu_1)
\frac{1}{2}(1+\sigma_2^3) K_+(\lambda_\alpha)
\sigma_\alpha^2 T(-\lambda_\alpha, \{ \nu \} \backslash \nu_1) \sigma_\alpha^2.
\end{align}
Next, we need the action of the operator $\mathcal{D}$ on the vectors 
created by $\mathcal{B}$ operators. Utilizing the following formula
\begin{align}
&\mathcal{D}(\lambda_i,\{ \nu \})
\mathcal{B}(\lambda_{i-1},\{ \nu \}) \cdots
\mathcal{B}(\lambda_{1},\{ \nu \})w_N^+ \nonumber \\
=&\sum_{k=1}^i \frac{\sh \eta \sh(2 \lambda_k+\eta)}{\sh 2 \lambda_k}
\Big\{ \frac{\sh(\lambda_k+\lambda_i) \sh(\lambda_k+\eta/2-\zeta_+)}
{\sh^2 \lambda_i-\sh^2(\lambda_k+\eta)} 
 \prod_{j=1}^N \frac{\sh(\lambda_k-\nu_j-\eta/2)}{\sh(\lambda_k-\nu_j+\eta/2)}
\prod_{\substack{j=1 \\ j \neq k}}^{i}
\frac{\sh^2 (\lambda_k+\eta)-\sh^2 \lambda_j}{\sh^2 \lambda_k-\sh^2 \lambda_j}
\nonumber \\
&+
\frac{\sh(\lambda_k-\lambda_i) \sh(-\lambda_k+\eta/2-\zeta_+)}
{\sh^2 \lambda_i-\sh^2(\lambda_k-\eta)} 
 \prod_{j=1}^N \frac{\sh(-\lambda_k-\nu_j-\eta/2)}
{\sh(-\lambda_k-\nu_j+\eta/2)}
\prod_{\substack{j=1 \\ j \neq k}}^{i}
\frac{\sh^2 (\lambda_k-\eta)-\sh^2 \lambda_j}{\sh^2 \lambda_k-\sh^2 \lambda_j}
\Big\} \nonumber \\
&\times 
\mathcal{B}(\lambda_{i},\{ \nu \}) \cdots
\check{\mathcal{B}}(\lambda_{k},\{ \nu \}) \cdots
\mathcal{B}(\lambda_{1},\{ \nu \}) w_N^+, \label{actionofd}
\end{align}
one has
\begin{align}
\psi_1(M,L)=&-c(-\lambda_M-\nu_1-\eta/2)
\prod_{j=1}^{M-1} \{b(-\lambda_j-\nu_1-\eta/2)
b(\lambda_j-\nu_1-\eta/2)  \}
\sum_{\beta=1}^M \frac{\sh \eta \sh(2 \lambda_\beta+\eta)}{\sh 2 \lambda_\beta}
\nonumber \\
&\times
\Big\{ \frac{\sh(\lambda_\beta+\lambda_M) \sh(\lambda_\beta+\eta/2-\zeta_+)}
{\sh^2 \lambda_M-\sh^2(\lambda_\beta+\eta)} 
 \prod_{j=2}^N \frac{\sh(\lambda_\beta-\nu_j-\eta/2)}
{\sh(\lambda_\beta-\nu_j+\eta/2)}
\prod_{\substack{j=1 \\ j \neq \beta}}^{M}
\frac{\sh^2 (\lambda_\beta+\eta)-\sh^2 \lambda_j}
{\sh^2 \lambda_\beta-\sh^2 \lambda_j}
\nonumber \\
&+
\frac{\sh(\lambda_\beta-\lambda_M) \sh(-\lambda_\beta+\eta/2-\zeta_+)}
{\sh^2 \lambda_M-\sh^2(\lambda_\beta-\eta)} 
 \prod_{j=2}^N \frac{\sh(-\lambda_\beta-\nu_j-\eta/2)}
{\sh(-\lambda_\beta-\nu_j+\eta/2)}
\prod_{\substack{j=1 \\ j \neq \beta}}^{M}
\frac{\sh^2 (\lambda_\beta-\eta)-\sh^2 \lambda_j}
{\sh^2 \lambda_\beta-\sh^2 \lambda_j}
\Big\} \nonumber \\
&\times w_{N-1}^-
\mathcal{B}(\lambda_N, \{ \nu \} \backslash \nu_1)
\cdots \mathcal{B}(\lambda_{L+1}, \{ \nu \} \backslash \nu_1)
\mathcal{G}(\lambda_L, \{ \nu \} \backslash \nu_1)
\nonumber \\
&\times \mathcal{B}(\lambda_{L-1}, \{ \nu \} \backslash \nu_1)
\cdots
\check{\mathcal{B}}(\lambda_{\beta}, \{ \nu \} \backslash \nu_1)
\cdots
\mathcal{B}(\lambda_1, \{ \nu \} \backslash \nu_1)
w_{N-1}^+, \label{intermediate}
\end{align}
getting rid of the operator $\mathcal{D}$.
We repeat the similar procedure we did as above to 
\eqref{intermediate}.
From the graphical description, one obtains
\begin{align}
\psi_1(M,L)=&-c(-\lambda_M-\nu_1-\eta/2)
\prod_{j=1}^{M-1} \{b(-\lambda_j-\nu_1-\eta/2)
b(\lambda_j-\nu_1-\eta/2)  \} b(-\lambda_L-\nu_2-\eta/2)
 \nonumber \\
&\times c(\lambda_L-\nu_2-\eta/2)
\sum_{\beta=1}^M
\prod_{\substack{j=1 \\ j \neq \beta }}^{L-1} \{b(-\lambda_j-\nu_2-\eta/2)
b(\lambda_j-\nu_2-\eta/2)  \}
\frac{\sh \eta \sh(2 \lambda_\beta+\eta)}{\sh 2 \lambda_\beta}
\nonumber \\
&\times
\Big\{ \frac{\sh(\lambda_\beta+\lambda_M) \sh(\lambda_\beta+\eta/2-\zeta_+)}
{\sh^2 \lambda_M-\sh^2(\lambda_\beta+\eta)} 
 \prod_{j=2}^N \frac{\sh(\lambda_\beta-\nu_j-\eta/2)}
{\sh(\lambda_\beta-\nu_j+\eta/2)}
\prod_{\substack{j=1 \\ j \neq \beta}}^{M}
\frac{\sh^2 (\lambda_\beta+\eta)-\sh^2 \lambda_j}
{\sh^2 \lambda_\beta-\sh^2 \lambda_j}
\nonumber \\
&+
\frac{\sh(\lambda_\beta-\lambda_M) \sh(-\lambda_\beta+\eta/2-\zeta_+)}
{\sh^2 \lambda_M-\sh^2(\lambda_\beta-\eta)} 
 \prod_{j=2}^N \frac{\sh(-\lambda_\beta-\nu_j-\eta/2)}
{\sh(-\lambda_\beta-\nu_j+\eta/2)}
\prod_{\substack{j=1 \\ j \neq \beta}}^{M}
\frac{\sh^2 (\lambda_\beta-\eta)-\sh^2 \lambda_j}
{\sh^2 \lambda_\beta-\sh^2 \lambda_j}
\Big\} \nonumber \\
&\times w_{N-2}^-
\mathcal{B}(\lambda_N, \{ \nu \} \backslash \{ \nu_1, \nu_2 \})
\cdots \mathcal{B}(\lambda_{L+1}, \{ \nu \} \backslash \{ \nu_1, \nu_2 \})
\mathcal{A}(\lambda_L, \{ \nu \} \backslash \{ \nu_1, \nu_2 \} )
\nonumber \\
&\times \mathcal{B}(\lambda_{L-1}, \{ \nu \} \backslash \{ \nu_1, \nu_2 \})
\cdots
\check{\mathcal{B}}(\lambda_{\beta}, \{ \nu \} \backslash \{ \nu_1, \nu_2 \})
\cdots
\mathcal{B}(\lambda_1, \{ \nu \} \backslash \{ \nu_1, \nu_2 \})
w_{N-2}^+.
\end{align}
Then, applying the formula
\begin{align}
&\mathcal{A}(\lambda_i,\{ \nu \})
\mathcal{B}(\lambda_{i-1},\{ \nu \}) \cdots
\mathcal{B}(\lambda_{1},\{ \nu \})w_N^+ \nonumber \\
=&\sum_{k=1}^i \frac{\sh \eta \sh(2 \lambda_k+\eta)}{\sh 2 \lambda_k}
\Big\{ \frac{ \sh(\lambda_k+\eta/2-\zeta_+)}
{\sh(\lambda_k+\lambda_i+\eta)} 
 \prod_{j=1}^N \frac{\sh(\lambda_k-\nu_j-\eta/2)}{\sh(\lambda_k-\nu_j+\eta/2)}
\prod_{\substack{j=1 \\ j \neq k}}^{i}
\frac{\sh^2 (\lambda_k+\eta)-\sh^2 \lambda_j}{\sh^2 \lambda_k-\sh^2 \lambda_j}
\nonumber \\
&+
\frac{\sh(-\lambda_k+\eta/2-\zeta_+)}
{\sh(\lambda_k-\lambda_i-\eta)} 
 \prod_{j=1}^N \frac{\sh(-\lambda_k-\nu_j-\eta/2)}
{\sh(-\lambda_k-\nu_j+\eta/2)}
\prod_{\substack{j=1 \\ j \neq k}}^{i}
\frac{\sh^2 (\lambda_k-\eta)-\sh^2 \lambda_j}{\sh^2 \lambda_k-\sh^2 \lambda_j}
\Big\} \nonumber \\
&\times 
\mathcal{B}(\lambda_{i},\{ \nu \}) \cdots
\check{\mathcal{B}}(\lambda_{k},\{ \nu \}) \cdots
\mathcal{B}(\lambda_{1},\{ \nu \}) w_N^+,
\end{align}
we have
\begin{align}
\psi_1(M,L)=&-c(-\lambda_M-\nu_1-\eta/2)
\prod_{j=1}^{M-1} \{b(-\lambda_j-\nu_1-\eta/2)
b(\lambda_j-\nu_1-\eta/2)  \} b(-\lambda_L-\nu_2-\eta/2)
 \nonumber \\
&\times c(\lambda_L-\nu_2-\eta/2)
\sum_{\beta=1}^M
\prod_{\substack{j=1 \\ j \neq \beta }}^{L-1} \{b(-\lambda_j-\nu_2-\eta/2)
b(\lambda_j-\nu_2-\eta/2)  \}
\frac{\sh \eta \sh(2 \lambda_\beta+\eta)}{\sh 2 \lambda_\beta}
\nonumber \\
&\times
\Big\{ \frac{\sh(\lambda_\beta+\lambda_M) \sh(\lambda_\beta+\eta/2-\zeta_+)}
{\sh^2 \lambda_M-\sh^2(\lambda_\beta+\eta)} 
 \prod_{j=2}^N \frac{\sh(\lambda_\beta-\nu_j-\eta/2)}
{\sh(\lambda_\beta-\nu_j+\eta/2)}
\prod_{\substack{j=1 \\ j \neq \beta}}^{M}
\frac{\sh^2 (\lambda_\beta+\eta)-\sh^2 \lambda_j}
{\sh^2 \lambda_\beta-\sh^2 \lambda_j}
\nonumber \\
&+
\frac{\sh(\lambda_\beta-\lambda_M) \sh(-\lambda_\beta+\eta/2-\zeta_+)}
{\sh^2 \lambda_M-\sh^2(\lambda_\beta-\eta)} 
 \prod_{j=2}^N \frac{\sh(-\lambda_\beta-\nu_j-\eta/2)}
{\sh(-\lambda_\beta-\nu_j+\eta/2)}
\prod_{\substack{j=1 \\ j \neq \beta}}^{M}
\frac{\sh^2 (\lambda_\beta-\eta)-\sh^2 \lambda_j}
{\sh^2 \lambda_\beta-\sh^2 \lambda_j}
\Big\} \nonumber \\
&\times \sum_{\substack{\alpha=1 \\ \alpha \neq \beta}}^L
\frac{\sh \eta \sh(2 \lambda_\alpha+\eta)}{\sh 2 \lambda_\alpha}
\Big\{ \frac{ \sh(\lambda_\alpha+\eta/2-\zeta_+)}
{\sh(\lambda_\alpha+\lambda_L+\eta)} 
 \prod_{j=3}^N \frac{\sh(\lambda_\alpha-\nu_j-\eta/2)}
{\sh(\lambda_\alpha-\nu_j+\eta/2)}
\prod_{\substack{j=1 \\ j \neq \alpha,\beta}}^{L}
\frac{\sh^2 (\lambda_\alpha+\eta)-\sh^2 \lambda_j}
{\sh^2 \lambda_\alpha-\sh^2 \lambda_j}
\nonumber \\
&+
\frac{\sh(-\lambda_\alpha+\eta/2-\zeta_+)}
{\sh(\lambda_\alpha-\lambda_L-\eta)} 
 \prod_{j=3}^N \frac{\sh(-\lambda_\alpha-\nu_j-\eta/2)}
{\sh(-\lambda_\alpha-\nu_j+\eta/2)}
\prod_{\substack{j=1 \\ j \neq \alpha, \beta}}^{L}
\frac{\sh^2 (\lambda_\alpha-\eta)-\sh^2 \lambda_j}
{\sh^2 \lambda_\alpha-\sh^2 \lambda_j}
\Big\} \nonumber \\
&\times Z_{2(N-2) \times (N-2)}( \{ \lambda \} \backslash 
\{ \lambda_\alpha, \lambda_\beta  \},
\{ \nu \} \backslash \{ \nu_1, \nu_2 \} ). \label{numerator}
\end{align}
Dividing \eqref{numerator} by the partition function
\eqref{boundarypartition} and simplifying, one gets
\begin{align}
&\Psi_1(M,L) \nonumber \\
=&\frac{\sh^2 \eta}{\det_N \chi(\{\lambda \}, \{\nu \})
\sh(\lambda_M+\nu_1-\eta/2) \sh(\lambda_L-\nu_2-\eta/2)
\prod_{j=1}^{M-1} \[ \sh^2(\nu_1-\eta/2)-\sh^2 \lambda_j  \]
} \nonumber \\
&\times
\frac{
\prod_{j=2}^N \[ \sh^2 \nu_1-\sh^2 \nu_j \]
\prod_{j=3}^N \[ \sh^2 \nu_2-\sh^2 \nu_j \]
}
{\prod_{j=M}^N \[ \sh^2(\nu_1+\eta/2)-\sh^2 \lambda_j  \]
\prod_{j=1}^L \[ \sh^2(\nu_2-\eta/2)-\sh^2 \lambda_j \]
\prod_{j=L+1}^N \[ \sh^2(\nu_2+\eta/2)-\sh^2 \lambda_j  \]
} \nonumber \\
&\times \sum_{\alpha=1}^L \sum_{\substack{\beta=1}}^M
(-1)^{\alpha+\beta} \epsilon_{\alpha \beta}
\det_{N-2} \chi(\{ \lambda \} \backslash 
\{ \lambda_\alpha, \lambda_\beta \}, \{ \nu \} \backslash
\{\nu_1, \nu_2  \}) \sum_{i,j=1}^2
F_{i,j}(\lambda_\alpha,\lambda_\beta),
\label{doublesum}
\end{align}
where
\begin{align}
\epsilon_{\alpha \beta}=\left\{
\begin{array}{cc}
1  & \alpha > \beta  \\
0  & \alpha = \beta \\
-1 & \alpha < \beta 
\end{array}
\right. ,
\end{align}
and
\begin{align}
F_{1,1}(\lambda_\alpha, \lambda_\beta)&=
\frac{G_1(\lambda_\alpha) H_1(\lambda_\beta)}
{\sh^2 (\lambda_\alpha+\eta)-\sh^2 \lambda_\beta},
\\
F_{1,2}(\lambda_\alpha, \lambda_\beta)&=
\frac{G_1(\lambda_\alpha) H_2(\lambda_\beta)}
{\sh^2 (\lambda_\alpha+\eta)-\sh^2 \lambda_\beta},
\\
F_{2,1}(\lambda_\alpha, \lambda_\beta)&=
\frac{G_2(\lambda_\alpha) H_1(\lambda_\beta)}
{\sh^2 (\lambda_\alpha-\eta)-\sh^2 \lambda_\beta},
\\
F_{2,2}(\lambda_\alpha, \lambda_\beta)&=
\frac{G_2(\lambda_\alpha) H_2(\lambda_\beta)}
{\sh^2 (\lambda_\alpha-\eta)-\sh^2 \lambda_\beta},
\end{align}
where $G_1(\lambda_\alpha), G_2(\lambda_\alpha), H_1(\lambda_\beta)$
and $H_2(\lambda_\beta)$ are
\begin{align}
G_1(\lambda_\alpha)=&\frac{\sh(\lambda_\alpha+\eta/2-\zeta_+)}
{\sh 2 \lambda_\alpha \sh(\lambda_\alpha+\lambda_L+\eta)}
\frac{\prod_{j=1}^L \[ \sh^2 (\lambda_\alpha+\eta)-\sh^2 \lambda_j  \]
\prod_{j=L+1}^N \[ \sh^2 \lambda_\alpha -\sh^2 \lambda_j \] 
}
{
\prod_{j=3}^N \[ \sh^2 \nu_j-\sh^2 (\lambda_\alpha+\eta/2) \]
}, \\
G_2(\lambda_\alpha)=&\frac{\sh(2 \lambda_\alpha+\eta)
\sh(\lambda_\alpha-\eta/2+\zeta_+)}
{\sh 2 \lambda_\alpha
\sh (2 \lambda_\alpha-\eta) \sh(\lambda_\alpha-\lambda_L-\eta)}
\frac{\prod_{j=1}^L \[ \sh^2 (\lambda_\alpha-\eta)-\sh^2 \lambda_j  \]
\prod_{j=L+1}^N \[ \sh^2 \lambda_\alpha -\sh^2 \lambda_j \] 
}
{
\prod_{j=3}^N \[ \sh^2 \nu_j-\sh^2 (\lambda_\alpha-\eta/2) \]
}, \\
H_1(\lambda_\beta)=&\frac{\sh(\lambda_\beta+\lambda_M)
\sh(\lambda_\beta+\eta/2-\zeta_+) \sh(\eta/2-\lambda_\beta-\nu_2)}
{\sh 2 \lambda_\beta \sh(\lambda_\beta+\nu_2+\eta/2)}
\nonumber \\
&\times \frac{
\prod_{j=1}^{M-1}\[ \sh^2(\lambda_\beta+\eta)-\sh^2 \lambda_j  \]
\prod_{j=M+1}^N \[ \sh^2 \lambda_\beta -\sh^2 \lambda_j \]
}
{\prod_{j=3}^N \[ \sh^2 \nu_j -\sh^2 (\lambda_\beta+\eta/2) \]},
\\
H_2(\lambda_\beta)=&\frac{\sh(2\lambda_\beta+\eta)
\sh(\lambda_\beta-\lambda_M)
\sh(-\lambda_\beta+\eta/2-\zeta_+) \sh(\lambda_\beta-\nu_2+\eta/2)}
{\sh 2 \lambda_\beta \sh (2 \lambda_\beta-\eta)
 \sh(\lambda_\beta-\nu_2-\eta/2)}
\nonumber \\
&\times \frac{
\prod_{j=1}^{M-1}\[ \sh^2(\lambda_\beta-\eta)-\sh^2 \lambda_j  \]
\prod_{j=M+1}^N \[ \sh^2 \lambda_\beta -\sh^2 \lambda_j \]
}
{\prod_{j=3}^N \[ \sh^2 \nu_j -\sh^2 (\lambda_\beta-\eta/2) \]}.
\end{align}
Let us set $\lambda_j$ as $\lambda_j=\lambda+z_j$.
Note that the sum in \eqref{doublesum} can be formally extended to $N$
since $G_1(\lambda_\alpha)=G_2(\lambda_\alpha)=0$ for $\alpha=L+1, \dots , N$
and $H_1(\lambda_\beta)=H_2(\lambda_\beta)=0$ for $\beta=M+1, \dots , N$.
Then, one finds
that \eqref{doublesum} can be expressed in the determinant form as
\begin{align}
&\Psi_1(M,L) \nonumber \\
=&\frac{\sh^2 \eta}{\det_N \chi(\{\lambda \}, \{\nu \})
\sh(\lambda_M+\nu_1-\eta/2) \sh(\lambda_L-\nu_2-\eta/2)
\prod_{j=1}^{M-1} \[ \sh^2(\nu_1-\eta/2)-\sh^2 \lambda_j  \]
} \nonumber \\
&\times
\frac{
\prod_{j=2}^N \[ \sh^2 \nu_1-\sh^2 \nu_j \]
\prod_{j=3}^N \[ \sh^2 \nu_2-\sh^2 \nu_j \]
}
{\prod_{j=M}^N \[ \sh^2(\nu_1+\eta/2)-\sh^2 \lambda_j  \]
\prod_{j=1}^L \[ \sh^2(\nu_2-\eta/2)-\sh^2 \lambda_j \]
\prod_{j=L+1}^N \[ \sh^2(\nu_2+\eta/2)-\sh^2 \lambda_j  \]
} \nonumber \\
&\times \det(\exp(z_j \partial_{\epsilon_1})|
\exp(z_j \partial_{\epsilon_2})|
\chi(\lambda+z_j,\nu_{k}))_{1 \le j \le N, 3 \le k \le N}
\sum_{i,j=1}^2 F_{i.j}(\lambda+\epsilon_1, \lambda+\epsilon_2)
|_{\epsilon_1=\epsilon_2=0}.
\end{align}
Now let us set $\nu_j=\nu+w_j$ and take the homogeneous limit
$\lambda_j \to \lambda, \nu_j \to \nu$
by putting $z_j, w_j , j=1, \dots, N$ to zero in the order 
$w_1, w_3, \dots w_N, w_2, z_1 \dots z_N$.
We have
\begin{align}
&\Psi_1(M,L)^{(h)} \nonumber \\
=&\frac{(N-1)! (N-2)! \sh^2 \eta \sh^{2N-3} 2 \nu}{\det_N \Phi
\[\sh^2 \nu-\sh^2 (\lambda-\eta/2)  \]
\[\sh^2 (\nu-\eta/2)-\sh^2 \lambda  \]^{L+M-1}
\[\sh^2 (\nu+\eta/2)-\sh^2 \lambda  \]^{2N-L-M+1}
} \nonumber \\
&\times \det(
\partial_{\epsilon_1}^{j-1}|
\partial_{\epsilon_2}^{j-1}|
\Phi_{j,k-2})_{1 \le j \le N, 3 \le k \le N}
\sum_{i,j=1}^2 F_{i.j}^{(h)}(\epsilon_1, \epsilon_2)
|_{\epsilon_1=\epsilon_2=0},
\end{align}
where
\begin{align}
\Phi_{j,k}=\partial_{\lambda}^{j-1}
\partial_{\nu}^{k-1} \chi(\lambda,\nu),
\end{align}
and
\begin{align}
F_{1,1}^{(h)}(\epsilon_1, \epsilon_2)&=
\frac{G_1^{(h)}(\epsilon_1) H_1^{(h)}(\epsilon_2)}
{\sh^2 (\lambda+\epsilon_1+\eta)-\sh^2 (\lambda+\epsilon_2)},
\\
F_{1,2}^{(h)}(\epsilon_1, \epsilon_2)&=
\frac{G_1^{(h)}(\epsilon_1) H_2^{(h)}(\epsilon_2)}
{\sh^2 (\lambda+\epsilon_1+\eta)-\sh^2 (\lambda+\epsilon_2)},
\\
F_{2,1}^{(h)}(\epsilon_1, \epsilon_2)&=
\frac{G_2^{(h)}(\epsilon_1) H_1^{(h)}(\epsilon_2)}
{\sh^2 (\lambda+\epsilon_1-\eta)-\sh^2 (\lambda+\epsilon_2)},
\\
F_{2,2}^{(h)}(\epsilon_1, \epsilon_2)&=
\frac{G_2^{(h)}(\epsilon_1) H_2^{(h)}(\epsilon_2)}
{\sh^2 (\lambda+\epsilon_1-\eta)-\sh^2 (\lambda+\epsilon_2)},
\end{align}
where $G_1^{(h)}(\epsilon_1), G_2^{(h)}(\epsilon_1),
H_1^{(h)}(\epsilon_2)$ and $H_2^{(h)}(\epsilon_2)$ are
\begin{align}
G_1^{(h)}(\epsilon_1)=&\frac{\sh(\lambda+\epsilon_1+\eta/2-\zeta_+)}
{\sh (2 \lambda+2 \epsilon_1) 
\sh(2 \lambda+\epsilon_1+\eta)}
\frac{
\[ \sh^2 (\lambda+\epsilon_1+\eta)-\sh^2 \lambda \]^L
\[ \sh^2 (\lambda+\epsilon_1) -\sh^2 \lambda \]^{N-L}
}
{
\[ \sh^2 \nu-\sh^2 (\lambda+\epsilon_1+\eta/2) \]^{N-2}
}, \\
G_2^{(h)}(\epsilon_1)=&
\frac{\sh(2 \lambda+2 \epsilon_1 +\eta)
\sh(\lambda+\epsilon_1-\eta/2+\zeta_+)}
{\sh (2 \lambda+2 \epsilon_1)
\sh (2 \lambda+2 \epsilon_1-\eta) \sh(\epsilon_1-\eta)}
\frac{
\[ \sh^2 (\lambda+\epsilon_1-\eta)-\sh^2 \lambda \]^L
\[ \sh^2 (\lambda+\epsilon_1) -\sh^2 \lambda \]^{N-L}
}
{
\[ \sh^2 \nu-\sh^2 (\lambda+\epsilon_1-\eta/2) \]^{N-2}
}, \\
H_1^{(h)}(\epsilon_2)=&\frac{\sh(2 \lambda+\epsilon_2)
\sh(\lambda+\epsilon_2 +\eta/2-\zeta_+) \sh(\eta/2-\lambda-\epsilon_2-\nu)}
{\sh (2 \lambda+2 \epsilon_2) \sh(\lambda+\epsilon_2+\nu+\eta/2)}
\nonumber \\
&\times \frac{
\[ \sh^2(\lambda+\epsilon_2+\eta)-\sh^2 \lambda  \]^{M-1}
\[ \sh^2 (\lambda+\epsilon_2) -\sh^2 \lambda \]^{N-M}
}
{\[ \sh^2 \nu -\sh^2 (\lambda+\epsilon_2+\eta/2) \]^{N-2}},
\\
H_2^{(h)}(\epsilon_2)=&\frac{\sh(2\lambda+2 \epsilon_2+\eta)
\sh \epsilon_2
\sh(-\lambda-\epsilon_2+\eta/2-\zeta_+) \sh(\lambda+\epsilon_2-\nu+\eta/2)}
{\sh (2 \lambda+2 \epsilon_2) \sh (2 \lambda+2 \epsilon_2-\eta)
 \sh(\lambda+\epsilon_2-\nu-\eta/2)}
\nonumber \\
&\times \frac{
\[ \sh^2(\lambda+\epsilon_2-\eta)-\sh^2 \lambda  \]^{M-1}
\[ \sh^2 (\lambda+\epsilon_2) -\sh^2 \lambda \]^{N-M}
}
{\[ \sh^2 \nu -\sh^2 (\lambda+\epsilon_2-\eta/2) \]^{N-2}}.
\end{align}

\section{Type $\mathrm{II}$}
\begin{figure}[htbp]
 \begin{center}
  \includegraphics[width=100mm]{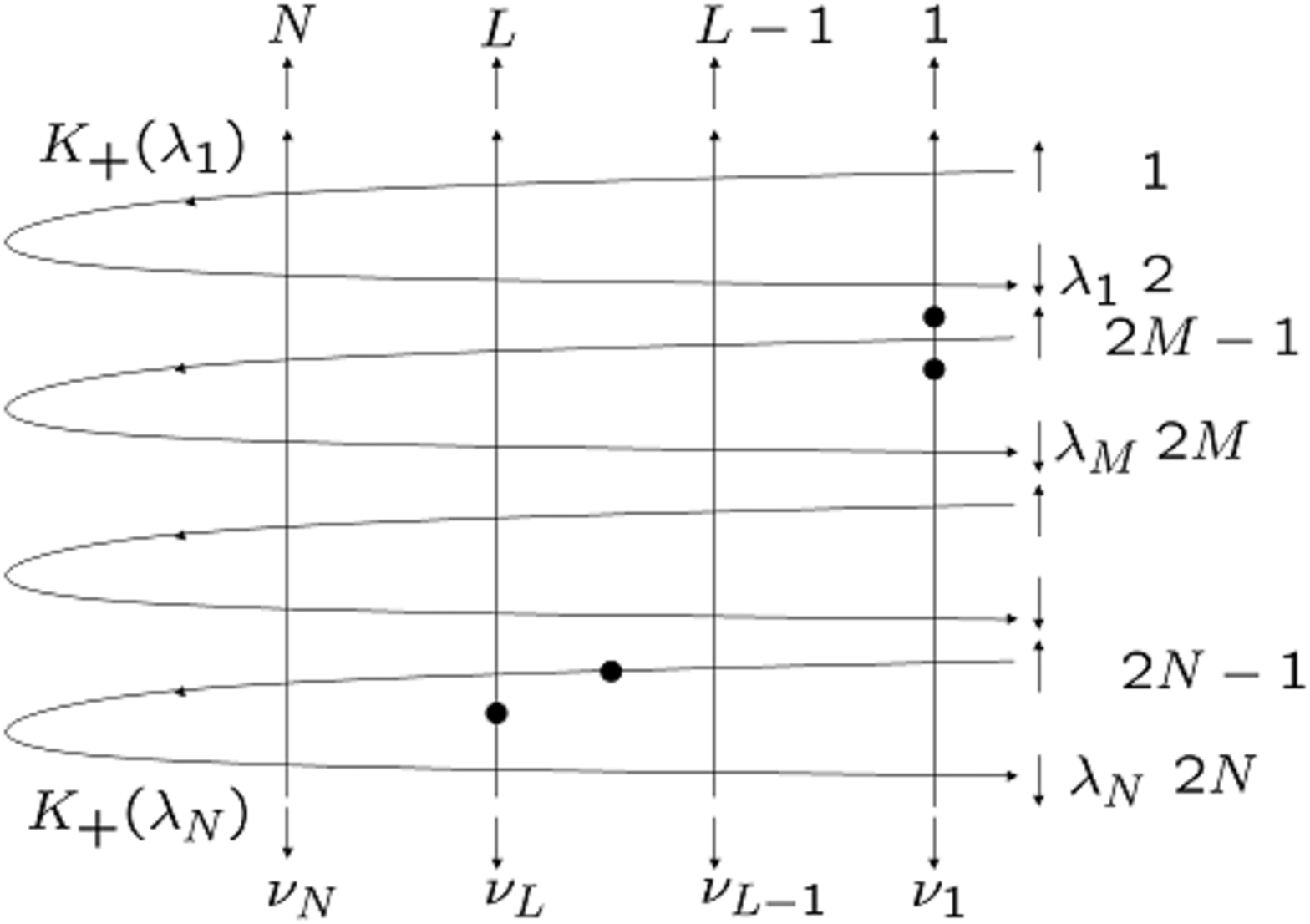}
 \end{center}
 \caption{Type $\mathrm{II}$ \eqref{type2}.}
 \label{fig:type2}
\end{figure}

In this section, we calculate another type
of two point function which we call Type $\mathrm{II}$
\begin{align}
\Psi_2(M,L)=\frac{\psi_2(M,L)}
{Z_{2N \times N}(\{ \lambda \}, \{ \nu \})},
\label{type2}
\end{align}
where
\begin{align}
\psi_2(M,L)=&w_N^- \mathcal{E}_L(\lambda_N, \{ \nu \})
\mathcal{B}(\lambda_{N-1}, \{ \nu \}) \cdots 
\mathcal{B}(\lambda_{M+1}, \{ \nu \}) \nonumber \\
&\times \mathcal{F}(\lambda_M, \{ \nu \})
\mathcal{B}(\lambda_{M-1}, \{ \nu \}) \cdots
\mathcal{B}(\lambda_1, \{ \nu \}) w_N^+,
\end{align}
\begin{align}
\mathcal{E}_L(\lambda_\alpha, \{ \nu \})
&=\downarrow_{2 \alpha} V_L(\lambda_\alpha, \{ \nu \})
\uparrow_{2 \alpha-1}, \\
V_L(\lambda_\alpha, \{ \nu \})&=T^{t_\alpha}(\lambda_\alpha, \{ \nu \})
 \frac{1}{2}(1+\sigma_L^3)
K_+(\lambda_\alpha) 
(\sigma_\alpha^2 L_{\alpha N}(-\lambda_\alpha, \nu_N) \sigma_\alpha^2) \cdots
(\sigma_\alpha^2 L_{\alpha L}(-\lambda_\alpha, \nu_L) \sigma_\alpha^2)
\nonumber \\
&\times
\frac{1}{2}(1+\sigma_\alpha^3)
(\sigma_\alpha^2 L_{\alpha L-1}(-\lambda_\alpha, \nu_{L-1})
\sigma_\alpha^2) \cdots
(\sigma_\alpha^2 L_{\alpha 1}(-\lambda_\alpha, \nu_1) \sigma_\alpha^2),
\end{align}
This two point function, depicted in Figure \ref{fig:type2},
gives the probability that the spin
on the first column is turned down just
on the $(2M-1)$-th row, and the spin on the $(2N)$-th row
row is turned down just on the $L$-th column.
Let us calculate this two point function. \\
First, with the help of the graphical description, we find
the numerator $\psi_2(M,L)$ becomes
\begin{align}
\psi_2(M,L)=&\sh(\lambda_N+\eta/2+\zeta_+)
b(-\lambda_N-\nu_L-\eta/2)c(\lambda_N-\nu_L-\eta/2)
\prod_{j=L+1}^N b(\lambda_N-\nu_j-\eta/2) \nonumber \\
&\times \langle 1, \cdots ,\check{L}, \cdots , N|
\mathcal{B}(\lambda_{N-1}, \{ \nu \}) \cdots
\mathcal{B}(\lambda_{M+1}, \{ \nu \}) \nonumber \\
&\times \mathcal{F}(\lambda_{M}, \{ \nu \})
\mathcal{B}(\lambda_{M-1}, \{ \nu \})
\cdots \mathcal{B}(\lambda_{1}, \{ \nu \})
 w_N^+ \nonumber \\
=&-\sh(\lambda_N+\eta/2+\zeta_+)
b(-\lambda_N-\nu_L-\eta/2)c(\lambda_N-\nu_L-\eta/2)
\prod_{j=L+1}^N b(\lambda_N-\nu_j-\eta/2) \nonumber \\
&\times c(\lambda_M-\nu_1-\eta/2)
\prod_{j=1}^{M-1} \{ b(-\lambda_j-\nu_1-\eta/2)b(\lambda_j-\nu_1-\eta/2)  \}
\nonumber \\
&\times
\langle 1, \cdots ,\check{L}, \cdots , N-1|
\mathcal{B}(\lambda_{N-1}, \{ \nu \} \backslash \nu_1) \cdots
\mathcal{B}(\lambda_{M+1}, \{ \nu \} \backslash \nu_1 ) \nonumber \\
&\times \mathcal{D}(\lambda_{M}, \{ \nu \} \backslash \nu_1)
\mathcal{B}(\lambda_{M-1}, \{ \nu \} \backslash \nu_1)
\cdots \mathcal{B}(\lambda_{1}, \{ \nu \} \backslash \nu_1)
 w_{N-1}^+, \label{intermediate2}
\end{align}
where $\langle 1, \cdots ,\check{L}, \cdots , n|=
\ \downarrow_1 \cdots \uparrow_{L} \cdots \downarrow_{n}$.
Next, utilizing \eqref{actionofd} and
\begin{align}
\prod_{j=1}^i \mathcal{B}(\lambda_j, \{ \nu \} ) w_N^+
=&\prod_{j=1}^i \frac{\sh(2 \lambda_j+\eta)}{\sh (2 \lambda_j)}
\sum_{\sigma_1=\pm} \cdots \sum_{\sigma_i=\pm}
\prod_{j=1}^i \{ (-\sigma_j) \sh(-\sigma_j
\lambda_j+\eta/2-\zeta_+) \} \nonumber \\
&\times
\prod_{j=1}^i \prod_{k=1}^N \frac{\sh (-\sigma_j \lambda_j-\nu_k-\eta/2)}
{\sh( -\sigma_j \lambda_j-\nu_k+\eta/2)}
\prod_{1 \le j < k \le i}
\frac{\sh(\sigma_j \lambda_j+\sigma_k \lambda_k-\eta)}
{\sh(\sigma_j \lambda_j+\sigma_k \lambda_k)}
\prod_{j=1}^i B(\sigma_j \lambda_j, \{ \nu  \}) w_N^+,
\end{align}
(see \cite{Wang2} for the rational case), 
one finds \eqref{intermediate2} can be expressed in
terms of one-row monodromy matrices as
\begin{align}
\psi_2(M,L)=&-\sh \eta \sh(\lambda_N+\eta/2+\zeta_+)
b(-\lambda_N-\nu_L-\eta/2)c(\lambda_N-\nu_L-\eta/2)
c(\lambda_M-\nu_1-\eta/2) \nonumber \\
&\times \prod_{j=L+1}^N b(\lambda_N-\nu_j-\eta/2)
\prod_{j=1}^{M-1} \{ b(-\lambda_j-\nu_1-\eta/2)b(\lambda_j-\nu_1-\eta/2)  \}
\prod_{j=1}^{N-1} \frac{\sh(2 \lambda_j+\eta)}{\sh( 2 \lambda_j)} \nonumber \\
&\times \sum_{\alpha=1}^M \sum_{\sigma_1=\pm} \cdots \sum_{\sigma_{N-1}=\pm}
\prod_{j=1}^{N-1} \{(-\sigma_j) \sh (-\sigma_j \lambda_j+\eta/2-\zeta_+) \}
\prod_{j=1}^{N-1} \prod_{k=2}^N 
\frac{\sh(-\sigma_j \lambda_j-\nu_k-\eta/2)}
{\sh(-\sigma_j \lambda_j -\nu_k+\eta/2)} \nonumber \\
&\times \frac{\sh(-\sigma_\alpha \lambda_\alpha+\lambda_M)}
{\sh^2 \lambda_M -\sh^2(-\sigma_\alpha \lambda_\alpha+\eta)}
\prod_{\substack{j=1 \\ j \neq \alpha }}^M
\frac{\sh^2( -\sigma_\alpha \lambda_\alpha+\eta)-\sh^2 \lambda_j}
{\sh^2 \lambda_\alpha -\sh^2 \lambda_j}
\prod_{ \substack{1 \le j < k \le N-1 \\ j,k \neq \alpha}}
\frac{\sh(\sigma_j \lambda_j+\sigma_k \lambda_k-\eta)}
{\sh(\sigma_j \lambda_j+\sigma_k \lambda_k)} \nonumber \\
&\times \langle 1 \cdots \check{L} \cdots N-1|
\prod_{\substack{j=1 \\ j \neq \alpha }}^{N-1}
B(\sigma_{j} \lambda_{j}, \{ \nu \} \backslash \nu_1) w_{N-1}^+.
\label{intermediate3}
\end{align}
We then change the
viewpoint to use the column monodromy matrix
\begin{align}
\overline{T}(\nu_j, \{ \lambda \})
=&L_{Nj}(\lambda_{N}, \nu_j)
\cdots L_{1j}(\lambda_{1}, \nu_j) \nonumber \\
=&\left(
\begin{array}{cc}
\overline{A}(\nu_j, \{ \lambda \}) & \overline{B}(\nu_j, \{ \lambda \}) \\
\overline{C}(\nu_j, \{ \lambda \}) & \overline{D}(\nu_j, \{ \lambda \})
\end{array}
\right),
\end{align}
instead of the row transfer matrix \eqref{onerow}.
Then one finds 
\begin{align}
&\langle 1 \cdots \check{L} \cdots N-1|
\prod_{\substack{j=1 \\ j \neq \alpha }}^{N-1}
B(\sigma_{j} \lambda_{j}, \{ \nu \} \backslash \nu_1) w_{N-1}^+
\nonumber \\
=&v_{N-2}^+ \overline{C}(\nu_N, \sigma (\{ \lambda \} 
\backslash \{ \lambda_\alpha, \lambda_N \}))
\cdots
\overline{C}(\nu_{L+1}, \sigma (\{ \lambda \} 
\backslash \{ \lambda_\alpha, \lambda_N \})) \nonumber \\
&\times
\overline{A}(\nu_L, \sigma (\{ \lambda \} 
\backslash \{ \lambda_\alpha, \lambda_N \})) 
\overline{C}(\nu_{L-1}, \sigma (\{ \lambda \} 
\backslash \{ \lambda_\alpha, \lambda_N \}))
\cdots
\overline{C}(\nu_2, \sigma (\{ \lambda \} 
\backslash \{ \lambda_\alpha, \lambda_N \}))
 v_{N-2}^- \nonumber \\
=&v_{N-2}^- \overline{B}(\nu_N, \sigma (\{ \lambda \} 
\backslash \{ \lambda_\alpha, \lambda_N \}))
\cdots
\overline{B}(\nu_{L+1}, \sigma (\{ \lambda \} 
\backslash \{ \lambda_\alpha, \lambda_N \})) \nonumber \\
&\times
\overline{D}(\nu_L, \sigma (\{ \lambda \} 
\backslash \{ \lambda_\alpha, \lambda_N \})) 
\overline{B}(\nu_{L-1}, \sigma (\{ \lambda \} 
\backslash \{ \lambda_\alpha, \lambda_N \}))
\cdots
\overline{B}(\nu_2, \sigma (\{ \lambda \} 
\backslash \{ \lambda_\alpha, \lambda_N \}))
 v_{N-2}^+
\label{change}
\end{align}
where
$v_{N-2}^+= \prod_{\alpha=1}^{N-2} \uparrow_\alpha$,
$v_{N-2}^-= \prod_{\alpha=1}^{N-2} \downarrow_\alpha$
and \\
$\sigma (\{ \lambda \} 
\backslash \{ \lambda_\alpha, \lambda_N \})
=\{\sigma_1 \lambda_1, \dots  , \sigma_{\alpha-1} \lambda_{\alpha-1},
\sigma_{\alpha+1} \lambda_{\alpha+1}, \dots ,
\sigma_{N-1} \lambda_{N-1}  \}$.
Utilizing
\begin{align}
&\overline{D}(\nu_i, \{ \lambda \})
\overline{B}(\nu_{i-1}, \{ \lambda \}) \cdots
\overline{B}(\nu_1, \{ \lambda \}) v_N^+ \nonumber \\
=&\sum_{k=1}^i \frac{\sh \eta}{\sh(\nu_i-\nu_k+\eta)}
\prod_{\substack{j=1 \\ j \neq k}}^i
\frac{\sh (\nu_j-\nu_k+\eta)}{\sh (\nu_j-\nu_k)}
\prod_{j=1}^N \frac{\sh(\lambda_j-\nu_k-\eta/2)}
{\sh(\lambda_j-\nu_k+\eta/2)} \nonumber \\
&\times
\overline{B}(\nu_i, \{ \lambda \})
\cdots
\check{\overline{B}}(\nu_{k}, \{ \lambda \}) \cdots
\overline{B}(\nu_1, \{ \lambda \}) v_N^+,
\end{align}
and the determinant representation of the 
partition function of the six vertex model on a $N \times N$
lattice with domain wall boundary condition
\begin{align}
v_N^- \prod_{j=1}^N \overline{B}(\nu_j, \{ \lambda \}) v_N^+
=\frac{\prod_{j,k=1}^N \sh(\lambda_j-\nu_k-\eta/2)
\det_N \phi(\{\lambda \},\{ \nu \})
}
{\prod_{1 \le j < k \le N} \sh(\nu_j-\nu_k)
\prod_{1 \le j < k \le n} \sh(\lambda_k-\lambda_j)},
\end{align}
where $\phi$ is an $N \times N$ matrix whose elements are
\begin{align}
\phi_{j k}&=\phi(\lambda_j, \nu_k), \\
\phi(\lambda, \nu)
&=\frac{\sh \eta}{\sh(\lambda-\nu+\eta/2)
\sh(\lambda-\nu-\eta/2)},
\end{align}
one can evaluate \eqref{change} as \cite{Motegi}
\begin{align}
&v_{N-2}^- \overline{B}(\nu_N, \sigma (\{ \lambda \} 
\backslash \{ \lambda_\alpha, \lambda_N \}))
\cdots
\overline{B}(\nu_{L+1}, \sigma (\{ \lambda \} 
\backslash \{ \lambda_\alpha, \lambda_N \})) \nonumber \\
&\times
\overline{D}(\nu_L, \sigma (\{ \lambda \} 
\backslash \{ \lambda_\alpha, \lambda_N \})) 
\overline{B}(\nu_{L-1}, \sigma (\{ \lambda \} 
\backslash \{ \lambda_\alpha, \lambda_N \}))
\cdots
\overline{B}(\nu_2, \sigma (\{ \lambda \} 
\backslash \{ \lambda_\alpha, \lambda_N \}))
 v_{N-2}^+
\nonumber \\
&=\frac{\prod_{\substack{j=1 \\ j \neq \alpha}}^{N-1} \prod_{k=2}^N
\sh(\sigma_j \lambda_j-\nu_k-\eta/2)
\det_{N-1} h(\{ \nu \} \backslash \nu_1, 
\sigma( \{ \lambda \} \backslash \{ \lambda_\alpha, \lambda_N \} ))}
{\prod_{2 \le j < k \le N} \sh(\nu_k-\nu_j)
\prod_{\substack{1 \le j < k \le N-1 \\ j,k \neq \alpha}} 
\sh(\sigma_j \lambda_j- \sigma_k \lambda_k)}.
\label{det}
\end{align}
Here, $h$ is an $(N-1) \times (N-1)$ matrix whose elements are given by
\begin{align}
h_{jk}=&\left\{
\begin{array}{ll}
\frac{\prod_{i=2}^{L-1} \sh(\nu_i-\nu_{k+1}+\eta)
\prod_{i=L+1}^N \sh(\nu_i-\nu_{k+1})}
{\prod_{i=1 , i \neq \alpha}^{N-1}
\sh(\sigma_i \lambda_i-\nu_{k+1}+\eta/2)},  & j=1, \\
\phi(\sigma_{j-1} \lambda_{j-1}, \nu_{k+1}), & j=2, \cdots , \alpha,  \\
\phi(\sigma_{j} \lambda_{j}, \nu_{k+1}), & j=\alpha+1, \cdots , N-1.  \\
\end{array}
\right.
\end{align}
Combining \eqref{boundarypartition}, \eqref{intermediate3}, \eqref{change}
and \eqref{det}, we can express $\Psi_2(M,L)$ using determinants as
\begin{align}
&\Psi_2(M,L) \nonumber \\
=&
\frac{\sh^2 \eta \sh(\lambda_N+\eta/2+\zeta_+) \sh(-\lambda_N-\nu_L-\eta/2)}
{\det_N \chi(\{\lambda \},\{ \nu \})
\sh(\lambda_M-\nu_1+\eta/2) \[ \sh^2(\nu_L-\eta/2)-\sh^2 \lambda_N \]}
\prod_{j=1}^{N-1} \frac{\sh(2 \lambda_j+\eta)}{\sh (2 \lambda_j)}
\nonumber \\
&\times \frac{\prod_{k=2}^N (\sh^2 \nu_k-\sh^2 \nu_1)
\prod_{2 \le j < k \le N}\sh(\nu_k+\nu_j)
\prod_{j=1}^{N-1} ( \sh^2 \lambda_j-\sh^2 \lambda_N )}
{\prod_{j=1}^{M-1} \[ \sh^2 (\nu_1-\eta/2)-\sh^2 \lambda_j \]
\prod_{j=M}^N  \[  \sh^2(\nu_1+\eta/2)-\sh^2 \lambda_j   \]
} \nonumber \\
&\times \frac{1}{\prod_{j=2}^L \[ \sh^2 (\nu_j+\eta/2)-\sh^2 \lambda_N  \]
\prod_{j=L+1}^N \[ \sh^2 \nu_j-\sh^2(\lambda_N+\eta/2)  \]} \nonumber \\
&\times \sum_{\alpha=1}^M \sum_{\sigma_1=\pm} \cdots \sum_{\sigma_{N-1}=\pm}
\prod_{j=1}^{N-1}\{(-\sigma_j) \sh(-\sigma_j \lambda_j+\eta/2-\zeta_+) \}
\frac{
\prod_{1 \le j < k \le N-1} \sh(\sigma_j \lambda_j+\sigma_k \lambda_k-\eta)
}
{
\prod_{j=1}^{N-1} \prod_{k=2}^N
\sh(-\sigma_j \lambda_j-\nu_k+\eta/2)
}
\nonumber \\
&\times 
\frac{\prod_{j=M+1}^{N-1} ( \sh^2 \lambda_\alpha -\sh^2 \lambda_j )
\prod_{j=1}^{M-1} \sh(\sigma_\alpha \lambda_\alpha-\sigma_j \lambda_j-\eta)
}{\prod_{j=2}^N \sh(\sigma_\alpha \lambda_\alpha-\nu_j-\eta/2)
\prod_{j=M}^{N-1} \sh(\sigma_\alpha \lambda_\alpha+\sigma_j \lambda_j-\eta) }
\nonumber \\
&\times (-1)^\alpha \sh(-\sigma_\alpha \lambda_\alpha+\lambda_M)
 \det_{N-1} h(\{ \nu \} \backslash \nu_1, 
\sigma( \{ \lambda \} \backslash \{ \lambda_\alpha, \lambda_N \} )).
\end{align}
\section{Conclusion}
In this paper, we considered two point functions
for the six vertex model on a $2N \times N$ lattice
with domain wall boundary condition and
left reflecting boundary. 
We calculated two types:
(Type $\mathrm{I}$)
the probability that the spin
on the first column is turned down just
on the $(2M-1)$-th row, and the spin on the second column
is turned down just on the $(2L)$-th row,
(Type $\mathrm{II}$)
the probability that the spin
on the first column is turned down just
on the $(2M-1)$-th row, and the spin on the $(2N)$-th
row is turned down just on the $L$-th column.
For Type $\mathrm{I}$, we could express it utilizing a single
determinant whose entries contain differential operators
which act on some functions.
For Type $\mathrm{II}$, reducing the double row monodromy matrices
to one row monodromy matrices and then changing the viewpoint
to use column monodromy matrices, we expressed it as combinations of
determinants.

It is interesting to extend the analysis to multipoint functions
such as the emptiness formation probability.
Another interesting direction is to study one point functions of
higher rank/spin generalized vertex models with domain wall
boundary condition. The thorough investigation of the off shell
structure \cite{MM,MM2}  should be crucial for the analysis.

\end{document}